\documentstyle{l-aa}
\begin{document}
\def\gtsima{$\; \buildrel > \over \sim \;$}
\def\simgt{\lower.5ex\hbox{\gtsima}}
\def\capo{\\ \indent}
\def\cali{\par\noindent}
\def\bequ{\begin{center}\begin{equation}}
\def\fequ{\end{equation}\end{center}}
\thesaurus{12(12.03.1; 11.17.3)}
\title{Ionization by early quasars and Cosmic Microwave Background anisotropies}
\author{N.~Aghanim\inst{1} \and F. X. D\'esert\inst{2} \and J. L. Puget\inst{1}
\and R. Gispert\inst{1}}
\offprints{N. Aghanim }
\institute{IAS-CNRS, Universit\'e Paris XI, Batiment 121, F-91405 Orsay 
Cedex
\and Laboratoire d'Astrophysique, Observatoire de Grenoble, BP 53, 
414 rue de la piscine, F-38041 Grenoble Cedex 9}
\date{Received date  / accepted date }
\maketitle
\begin{abstract}
\end{abstract}
The y axis of figure 9, which gives the power amplitude ($l(l+1)C_l$) in units 
of $10^{-12}$, is incorrectly 
reported. The correct values are obtained by dividing the amplitudes by 16. 
All the relative values remain valid, and this therefore does not change the
conclusions of the paper.
\begin{acknowledgements}
We thank A. Gruzinov \& W. Hu for bringing this to our attention.
\end{acknowledgements}

%
%
\end{document}